\documentclass[10pt]{article}
\usepackage[letterpaper]{geometry}
\usepackage{hicss51}
\usepackage{times}
\usepackage[none]{hyphenat}
\usepackage{url}
\usepackage{latexsym}
\usepackage{indentfirst}

\usepackage{cite}

\usepackage[shortlabels]{enumitem}
\usepackage{csquotes}

  \usepackage[pdftex]{graphicx}
  \graphicspath{{pdf/}{png/}}
  \DeclareGraphicsExtensions{.pdf,.jpeg,.png}
\makeatletter
\renewcommand*{\thetable}{\arabic{table}}
\renewcommand*{\thefigure}{\arabic{figure}}
\let\c@table\c@figure
\makeatother 

\makeatletter
    \renewcommand{\fps@figure}{tbp}         
\makeatother

\usepackage{float}
\usepackage{tikz}
  \usetikzlibrary{matrix,arrows,decorations.pathmorphing}

\definecolor{gray1}{HTML}{A4A4A4}
\definecolor{gray2}{HTML}{585858}
\definecolor{gray3}{HTML}{2E2E2E}
\definecolor{gray4}{HTML}{FAFAFA}

\tikzstyle{decision} = [diamond, draw, fill=blue!20,
    text width=4.5em, text badly centered, node distance=2.5cm, inner sep=0pt]
\tikzstyle{block} = [rectangle, draw, fill=gray1,
    text width=5em, text centered, rounded corners, minimum height=4em]
\tikzstyle{line} = [draw, very thick, color=black!50, -latex']
\tikzstyle{cloud} = [draw, rectangle,fill=gray4, node distance=2.5cm,
    minimum height=2em, text centered, text width=7em]

\tikzstyle{start}=[shape=circle,double,double distance=1pt,draw=black!50,fill=gray4,minimum size=2cm,inner sep=0pt]
\tikzstyle{state}=[shape=circle,draw=black!50,fill=gray1,minimum size=2.5cm,inner sep=0pt]
\tikzstyle{observation}=[shape=rectangle,draw=black!50,fill=gray4,minimum width=1.5cm,minimum height=1cm,inner sep=1pt]
\tikzstyle{lightedge}=[<-,dotted]
\tikzstyle{mainstate}=[state,thick]
\tikzstyle{mainedge}=[<-,thick]

\usepackage[cmex10]{amsmath}
\usepackage[amsmath,hyperref,framed]{ntheorem} 
  \theoremstyle{plain}
     
  \theoremstyle{nonumberplain}
    \theoremseparator{.}
    \theoremheaderfont{\bfseries}
    \theorembodyfont{\normalfont}
    \theoremsymbol{$\blacksquare$}
      \RequirePackage{amssymb}

      \makeatletter
    \let\copy@theorem@headerfont=\theorem@headerfont
    \newcommand{\my@theorem@headerfont}{%
        \boldmath\copy@theorem@headerfont\unboldmath
      }
    \let\theorem@headerfont=\my@theorem@headerfont
      \makeatother

\theoremstyle{nonumberplain}
\theoremseparator{.}
\usepackage{enumitem} 
  \setlist{itemsep=1ex plus0.2ex, leftmargin=*, align=left}

\makeatletter
\newcommand{\labitem}[2]{%
\def\@itemlabel{\textbf{#1}}
\item
\def\@currentlabel{#1}\label{#2}}
\makeatother

\makeatletter
\newcommand{\headingitem}[1]{%
\vspace{0.3cm}
\def\@itemlabel{\textbf{#1}}
\item
\def\@currentlabel{#1}
\addtocounter{enumi}{-1}
}
\makeatother

\usepackage{subfig}

\usepackage{url}

\usepackage{xspace}
\newcommand\ie{i.\,e.\xspace}
\newcommand\eg{e.\,g.\xspace}

\usepackage{siunitx}
\usepackage{tabularx}
\usepackage{csvsimple}
\usepackage{csquotes}
\usepackage{multirow}
\usepackage{lipsum}

\usepackage[
    colorinlistoftodos,
    textsize=footnotesize,
        ]{todonotes}

\usepackage{booktabs}
\usepackage{dcolumn}

\makeatletter
\renewcommand{\fps@figure}{H}         
\renewcommand{\fps@table}{H}         
\makeatother 

\usepackage{multirow}
\usepackage{tikz}
\usetikzlibrary{matrix,arrows,decorations.pathmorphing}

\definecolor{gray1}{HTML}{A4A4A4}
\definecolor{gray2}{HTML}{757575}
\definecolor{gray3}{HTML}{FAFAFA}
\definecolor{gray4}{HTML}{F2F2F2}

\usepackage{soul}
\usepackage{xcolor}

\newcommand{\ctextdarkgray}[1]{%
  \begingroup
  \sethlcolor{gray1}%
  \hl{#1}%
  \endgroup
}

\newcommand{\ctextlightgray}[1]{%
  \begingroup
  \sethlcolor{gray4}%
  \hl{#1}%
  \endgroup
}
\usepackage{pgfplots}
\pgfplotsset{compat=newest}
\usepackage{pgfplotstable}
\pgfplotsset{%
	legend style={font=\small},
	label style={font=\small},
	tick label style={font=\small},
	no markers,
	width=\linewidth,
	legend cell align=left,
	every axis/.append style={line width=0.75pt},
}

\usepackage{collcell}

\usepackage{ragged2e}
  
\newcommand{\PreserveBackslash}[1]{\let\temp=\\#1\let\\=\temp}
\newcolumntype{v}[1]{>{\PreserveBackslash\RaggedRight\hspace{0pt}}p{#1}}

  \usepackage{adjustbox}
\newcolumntype{Q}[2]{%
    >{\adjustbox{angle=#1,lap=\width-(#2)}\bgroup}%
    l%
    <{\egroup}%
}

\usepackage[%
  sort&compress,
]{cleveref}
  \crefname{subsection}{section}{sections}
  \Crefname{subsection}{Section}{Sections}
  \crefname{lemma}{lemma}{lemmas}
  \Crefname{lemma}{Lemma}{Lemmas}  
  \crefname{equation}{equation}{equations}  
  \Crefname{equation}{Equation}{Equations}  
  \crefname{proposition}{proposition}{propositions}
  \Crefname{proposition}{Proposition}{Propositions}
  \crefname{corollary}{corollary}{corollaries}
  \Crefname{corollary}{Corollary}{Corollaries}
  \crefname{remark}{remark}{remarks}
  \Crefname{remark}{Remark}{Remarks}
  \crefname{definition}{definition}{definitions}
  \Crefname{definition}{Definition}{Definitions}
  \crefname{example}{example}{examples}
  \Crefname{example}{Example}{Examples}
  \crefname{conjecture}{conjecture}{conjectures}
  \Crefname{conjecture}{Conjecture}{Conjectures}
  \crefname{proof}{proof}{proofs}
  \Crefname{proof}{Proof}{Proofs}
  \crefname{enumi}{rule}{rules}
  \Crefname{enumi}{Rule}{Rules}

\usepackage{arydshln}
\newcommand{\NEUTRAL}{\multicolumn{1}{c}{--}}

\title{Sentence-Level Sentiment Analysis of Financial News Using Distributed Text Representations and Multi-Instance Learning}

\author{Bernhard Lutz \\
  University of Freiburg, Germany \\
  {\underline{ bernhard.lutz@is.uni-freiburg.de}} \\\And
  Nicolas Pr\"ollochs \\
  University of Oxford, UK \\
  {\underline{ nicolas.prollochs@eng.ox.ac.uk} }\\\And 
  Dirk Neumann \\
  University of Freiburg, Germany \\
  {\underline{dirk.neumann@is.uni-freiburg.de}} \\}

\date{}

\begin{document}
\maketitle

\begin{abstract}
Researchers and financial professionals require robust computerized tools that allow users to rapidly operationalize and assess the semantic textual content in financial news. However, existing methods commonly work at the document-level while deeper insights into the actual structure and the sentiment of individual sentences remain blurred. As a result, investors are required to apply the utmost attention and detailed, domain-specific knowledge in order to assess the information on a fine-grained basis.
To facilitate this manual process, this paper proposes the use of distributed text representations and multi-instance learning to transfer information from the document-level to the sentence-level. Compared to alternative approaches, this method features superior predictive performance while preserving context and interpretability. Our analysis of a manually-labeled dataset yields a predictive accuracy of up to \SI{69.90}{\percent}, exceeding the performance of alternative approaches by at least \num{3.80} percentage points. Accordingly, this study not only benefits investors with regard to their financial decision-making, but also helps companies to communicate their messages as intended. 
\end{abstract}


%

\section{Introduction}
\label{sec:intro}
\raggedbottom

Companies around the world are required by law to publish information that has the potential to influence their valuation \cite{Benston.1973}. 
These financial news releases serve as an important source of information for investors considering exercising ownership in stock, as they trigger subsequent movements in stock prices \cite{Kearney.2014, Loughran.2016}. 
Besides quantitative numbers, such as sales volume or earnings forecasts, financial news also contain a substantial amount of qualitative content. Although this textual information is more difficult to assess, it is still relevant to the valuation of a company \cite{Tetlock.2008, Loughran.2011}. Hence, investors are required to carefully evaluate language and word choice in financial news and then decide whether to exercise ownership in the stock in question~\cite{Carter.1999}. 


Due to the sheer amount of available financial information, it is of great importance for financial professionals to possess computerized tools to operationalize the textual content of financial news. Over the last several years, researchers have created a great number of decision support systems that process financial news; in order to predict the resulting stock market reaction. The overwhelming majority of such systems described in previous works consider every financial news item as a single document with a given label, \ie the stock market reaction (\eg~\cite{Alfano.2018, Prollochs.2016b, Schumaker.2012}). For the purpose of text categorization, researchers then transform documents into a representation suitable for the learning algorithm and the classification task. The usual method of feature extraction is the \textit{bag-of-words} approach \cite{Prollochs.2015}, which treats each document as a large and sparse vector that counts the frequency of a given set of terms, or $n$-grams. Although existing studies in this direction have produced remarkably robust results, the bag-of-words approach comes with multiple drawbacks, such as missing negation context and information loss. For instance, in the sentence \emph{\enquote{The company reduced its costs and increased its profit margin}}, the bag-of-words approach is unable to distinguish the meaning of words in this arrangement from a sentence with a slightly different word order, such as an exchange of \emph{\enquote{costs}} and \emph{\enquote{profit}}. 


Apart from the general difficulty of predicting future stock market returns, previous approaches suffer from further methodological challenges that reduce their helpfulness for researchers and practitioners. As a primary drawback, they typically work at the document-level, while deeper insights into the actual structure and polarity of individual sentences remain unavailable. However, financial news typically entail more than one aspect and thus, different sentences in a single text are likely to express different sentiments \cite{Lerman.2010}. This limitation not only hampers a fine-grained study of financial news, but also shows that \emph{\enquote{state-of-the-art sentiment analysis methods' sentiment polarity classification performances are subpar, which affects the sentiment-related analysis and conclusions drawn from it}} \cite{Abbasi.2016}. 
As a result, investors are still required to apply the utmost attention and detailed, domain-specific knowledge in order to assess the information on a fine-grained basis. In the same vein, companies and investor relations departments are lacking a decision support tool to assist them in communicating their message as intended. 


Hence, the purpose of this paper is to compare methods for operationalizing the textual content of financial news on a fine-grained basis. As a main contribution, we thereby propose a novel method that allows one to assess the semantic orientation of individual sentences and text fragments in financial news. To accomplish this task, we use a two-step approach. First, \emph{distributed text representations} allow for the preservation of the context-dependent nature of language, thereby overcoming some of the shortcomings of the bag-of-words approach. Second, \emph{multi-instance learning} allows one to train a classifier that can be used to transfer information from the document-level to the sentence-level \cite{Kotzias.2015}. In our scenario, a document is represented by a financial news item, whereas the document label is represented by the reaction of investors on the stock market. Based on this information, our approach learns polarity labels for the individual sentences within the financial document. In a nutshell, the combination of distributed text representations and multi-instance learning allows similar sentences to be classified with the same polarity label and differing sentences with the opposite polarity label. Our later analysis shows that this approach yields superior predictive performance and does not require any kind of manual labeling, as it is solely trained on the market reaction following the publication of a news item.


Our study immediately suggests manifold implications for researchers and practitioners. Financial professionals and investors can benefit from our tool, which allows them to easily distinguish between positive and negative text fragments in financial news based on statistical rigor. In contrast to existing approaches that merely predict the stock market reaction in response to financial news on a document-level, our method infers the individual aspects that are expressed in different sentences. This mitigates the risk of human investors being outperformed by automated traders and allows users to place orders in a shorter time \cite{GroKlumann.2011}. Based on this, company executives and investor relations departments may wish to consider choosing their language strategically so as to ensure that their message is interpreted as intended.  


The remainder of this work is structured as follows. In \Cref{sec:relatedwork}, we provide an overview of literature that performs sentiment analysis of financial news. In addition, we highlight the drawbacks of current approaches with regard to studying sentiment on a fine-grained level. Subsequently, \Cref{sec:datamodel} introduces our data sources and the way in which we integrate distributed text representations and multi-instance learning to infer sentence labels for financial news. \Cref{sec:results} presents our results, while 
\Cref{sec:discussion} discusses the implications of our study for researchers and practitioners. \Cref{sec:conclusion} concludes.

\section{Background}
\label{sec:relatedwork}


A tremendous amount of literature has examined the extent to which stock market prices are correlated with the information provided in financial news. While early studies have established a robust link between quantitative information in financial disclosures and stock market returns, researchers nowadays have \emph{\enquote{intensified their efforts to understand how sentiment impacts on individual decision-makers, institutions and markets}} \cite{Kearney.2014}. In the existing literature, sentiment is predominately considered a measure of the qualitative information in financial news, referring to the degree of positivity or negativity of opinions shared by the authors with regard to individual stocks or the overall market \cite{Loughran.2016, Kearney.2014}. In this context, the overwhelming majority of studies uses bag-of-words approaches to explain stock market returns, \eg by the linguistic tone of ad~hoc announcements (\eg~\cite{Prollochs.2016b}), 8-K filings (\eg~\cite{Hannemann.2018, Prollochs.2018}), newspaper articles (\eg~\cite{Tetlock.2007}) or company press releases~(\eg~\cite{Henry.2008}). Comprehensive literature overviews regarding textual sentiment analysis of financial news can be found in \cite{Nassirtoussi.2014}, as well as \cite{Loughran.2016}. 


Although many decision support systems have already been created for the prediction of the stock market direction, \eg \cite{Chiong.2018, Kraus.2017, Prollochs.2016b, Schumaker.2012}, their performance still remains unsatisfactory \cite{Nassirtoussi.2014} and only marginally better than random guessing. One possible explanation is that determining the sentiment only at the document-level does not account for the relevances of different text segments. Different sentences in financial news releases typically focus on different aspects and express different sentiments \cite{Lerman.2010}. Hence, an accurate classification of sentences would allow researchers not only to improve existing prediction systems but also to perform more fine-grained explanatory analyses on financial news.

Since the aforementioned hurdles limit the degree of severity of sentiment analysis applications, studies that analyze financial news on a fine-grained level are rare. As one of very few examples, \cite{Allee.2015} use a dictionary-based approach based on the Loughran-McDonald finance-specific dictionary to study the role of sentiment dispersion in corporate communication. The authors find that the distribution of sentiment is closely associated with investors' reactions to the textual narratives. 
Another study \cite{Li.2010b} acknowledges the drawbacks of dictionary-based methods and instead uses a Na\"ive Bayes approach to train a sentence classifier based on a set of \num{30000} manually-labeled sentences drawn from the forward-looking statements found in the Management Discussion and Analysis section of 10-K filings. However, apart from the fact that assigned manual labels are highly subjective, the utilized methodology suffers from the bag-of-words disadvantages, such as missing context and information loss \cite{Nassirtoussi.2014}.
Recently, SemEval-2017 conducted a challenge called \textit{Fine-Grained Sentiment Analysis on Financial Microblogs and News} \cite{Cortis.2017}. The task was to predict individual sentiment scores for companies/stocks mentioned in financial microblogs. The proposed methods utilize manually-labeled text segments in combination with supervised learning for text classification. Yet, the resulting prediction models are highly domain-specific and not easily generalizable to alternative text sources.  





Hence, this paper addresses the following research goal: we compare and propose algorithms to predict the sentiment of individual sentences in financial news. As a remedy for the drawbacks of previous approaches, we later devise a more fine-grained approach based on distributed text representations and multi-instance learning that allows for the transfer of information from the document-level to the sentence-level. Although multi-instance learning has been successfully applied for several machine learning tasks \cite{Carbonneau.2018}, including image categorization, text categorization, face detection and computer-aided medical diagnosis \cite{Dundar.2007}, we are not aware of any publication that utilizes this method to infer sentence labels for financial news. Moreover, to the best of our knowledge, this is the first study that compares methods for sentence-level sentiment analysis of financial news.

\section{Materials and Methods}
\label{sec:datamodel}

\begin{figure}
\centering
\includegraphics[width = \linewidth]{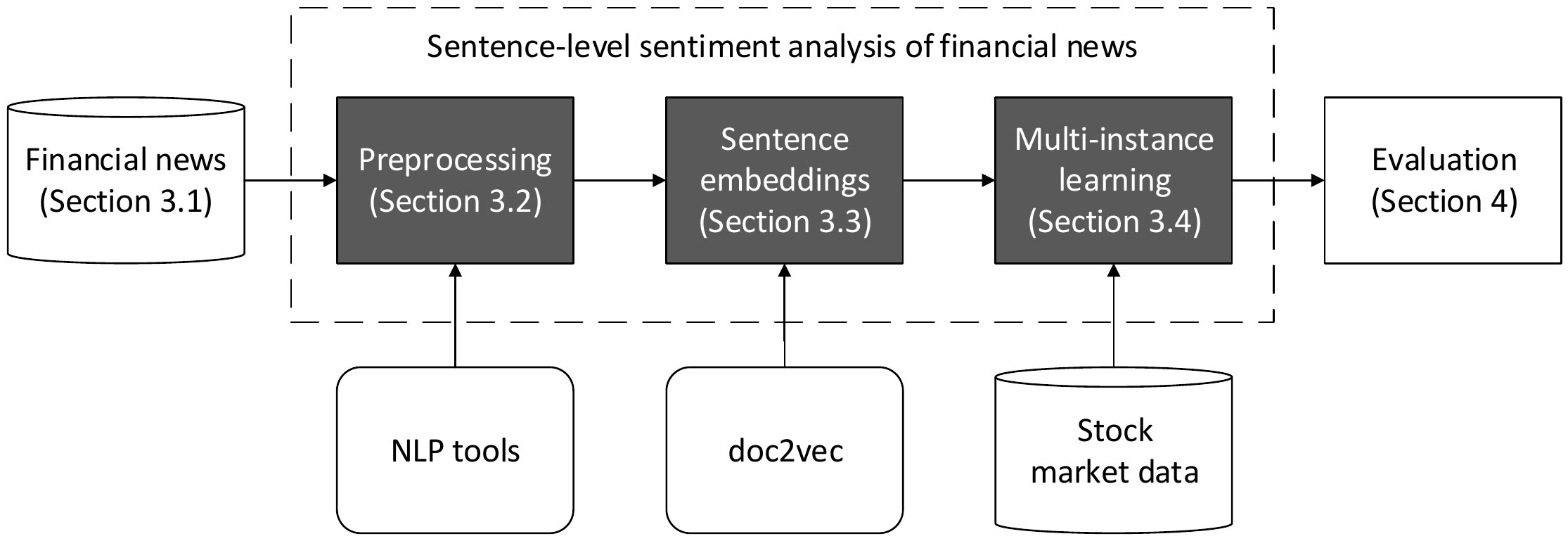}
\caption{Research model for sentence-level sentiment analysis of financial news.}
\label{fig:concept}
\end{figure}

In this section, we introduce our dataset and present our method for studying financial news at the sentence-level. \Cref{fig:concept} presents our research methodology. In a first step, we perform several preprocessing operations using tools from natural language processing. Second, the textual data is mapped to a vector-based representation using sentence embeddings. Third, we combine the vector representations with the historic stock market returns of companies to train a sentence-level classifier using multi-instance learning. The method is thoroughly evaluated and compared to alternative approaches in \Cref{sec:results}. 

\subsection{Dataset}


Our financial news dataset consists of \num{9502} German regulated ad~hoc announcements\footnote{Kindly provided by Deutsche Gesellschaft f{\"u}r Ad-Hoc-Publizit{\"a}t~(DGAP).} from between January 2001 and September 2017. As a requirement, each ad~hoc announcement must contain at least 50 words and be written in English. Companies in our dataset have published as few as \num{1} ad~hoc announcement, but also as many as \num{153}, with a median number of \num{10} announcements per company. The average number of ad~hoc announcements published per month is \num{46.80} during our period of study. The mean length of a single ad~hoc announcement is \num{508.98} words or \num{18.21} sentences. The average length of a sentence in our dataset is \num{28.89} words.
In research, ad~hoc announcements are a frequent choice~(\eg~\cite{Chiong.2018, Groth.2011, Hagenau.2013, Prollochs.2018b}) when it comes to evaluating and comparing methods for sentiment analysis. Additionally, this type of news corpus presents several advantages: ad~hoc announcements must be authorized by company executives, the content is quality-checked by the Federal Financial Supervisory Authority, 
and several publications confirm their relevance to the stock market (\eg~\cite{Prollochs.2016b}). 


In order to study the stock market reaction, we use the daily \emph{abnormal return} of the company that has published the financial item in question. For this purpose, we use the common event study methodology \cite{MacKinlay.1997}, whereby we determine the normal return, \ie the return which is expected in the absence of a news disclosure, with the help of a market model. This market model assumes a stable linear relation between market return and normal return. Concordant with the related literature, we model the market return using a stock market index, namely, the CDAX, along with an event window of 30~trading days prior to the news disclosure.  Finally, we determine the abnormal return as the difference between actual and normal returns. Here, all financial market data originates from Bloomberg.  

\subsection{Preprocessing}


We apply several common filtering steps to our dataset, which allows us to reduce the effect of confounding influences in our later analysis. Concordant with the related literature, we account for extreme stock price effects by removing penny stocks with a price lower than $\$1$ and by omitting outliers at the \SI{1}{\percent} level \cite{Zhang.2012}. In addition, we remove ad~hoc announcements for which we were not able to determine the stock market reaction from Bloomberg. These filtering steps result in a sample of \num{6360} ad~hoc announcements.


Next, we perform several common preprocessing steps on the textual data, in order to remove formatting and noisy content. First, by using a list of cut-off patterns, we omit contact addresses and HTML formatting. Second, we convert each ad~hoc announcement to lower case and replace dates, positive and negative numbers, and URLs with appropriate tokens. Third, we tokenize infrequent terms that appear fewer than five times \cite{Kotzias.2015}. These preprocessing steps reduce the size of the vocabulary from \num{34910} words to \num{10969} words.


Finally, we use the sentence-splitting tool from Stanford CoreNLP \cite{Manning.2014} to partition each ad~hoc announcement into sentences. It is worth noting that this approach also addresses the frequently-found challenges in previous works regarding the accurate division of financial items into sentences because \emph{\enquote{the presence of extensive lists, technical terminology, and other formatting complexities, makes sentence disambiguation especially challenging in accounting disclosures}}~\cite{Loughran.2016}.
 We observe that \SI{93.76}{\percent} of all ad~hoc announcements contain between \num{5} and \num{40} sentences, while a few ad~hoc announcements are of very short or excessive length. Thus, to ensure comparability, we remove all ad~hoc announcements with lengths in the highest and lowest percentiles from our dataset. Our final corpus consists of \num{6258} ad~hoc announcements. 
The total number of sentences across all ad~hoc announcements is \num{91315}. 
Out of all disclosures, a total number of \num{3486} ad~hoc announcements (\SI{55.70}{\percent}) resulted in a positive abnormal return, whereas \num{2772} (\SI{44.30}{\percent}) led to a negative abnormal return.

\subsection{Distributed Text Representations} 


The accuracy of sentiment analysis depends heavily on the representation of the textual data and the selection of features \cite{Schumaker.2012}. To overcome the drawbacks of the frequently employed bag-of-words approach, such as missing context and information loss, we take advantage of recent advances in learning distributed representations for text. 


For this purpose, we employ the \textit{doc2vec} library developed by Google \cite{Le.2014}. This library is based on a deep learning model that creates numerical representation for texts, regardless of their length. Specifically, the underlying model allows one to create distributed representations of sentences and documents by mapping the textual data onto a vector space. 

The word vectors being used in this model have several useful properties. First, more similar words are mapped to more similar vectors. For instance, the word \emph{cost} is mapped closer to \emph{debt} than to \emph{company}. Second, the feature vectors also fulfill simple algebraic properties such as, for example, \emph{king} - \emph{man} + \emph{woman} = \emph{queen}. Thus, in contrast to the bag-of-words approach, the \emph{doc2vec} library incorporates context-specific information and semantic similarities. As a further advantage, the feature space of the sentence representations is typically in a relatively small range between \num{200} and \num{400} dimensions (as compared to the several thousand often found with bag-of-words models). The feature representations created by the \textit{doc2vec} library have been shown to significantly increase the predictive performance of machine learning models for text classification \cite{Le.2014}. 


For the training of our \emph{doc2vec} model, we initialize the word vectors with the vectors from the pre-trained Google News dataset\footnote{Available from the Google code archive at \url{https://code.google.com/archive/p/word2vec/}.}, which is the predominant choice in the previous literature (\eg~\cite{Tang.2014b
}). Here, we use the hyperparameter settings developed by \cite{Lau.2016} during an extensive analysis. Subsequently, we generate vector representations for all sentences in our sample. These sentence embeddings are used in the next section as input data to train a sentence-level classifier using multi-instance learning. 

\subsection{Sentence-Level Sentiment Analysis Using Multi-Instance Learning}


We are facing a problem in which the observations (documents) contain groups of instances (sentences) instead of a single feature vector, where each group is associated with a label (stock returns). Formally, let $X = \{\boldsymbol{x}_i\}, i=1\dots N$ denote the set of all instances in all groups, $N$ the number of instances, $D$ the set of groups and $K$ the number of groups. Each group $D_k=(\mathcal{G}_k, l_k)$ consists of a multiset of instances $\mathcal{G}_k \subseteq X$ and is assigned a label $l_k$ ($0$ for negative and $1$ for positive). The learning task is to train a classifier $y$ with parameters $\boldsymbol{\theta}$ to infer instance labels $y_{\boldsymbol{\theta}}(\boldsymbol{x}_i)$ given only the group labels.

The above problem is a multi-instance learning problem \cite{Dietterich.1997} which can be solved by constructing a loss function consisting of two components: (a)~a term that punishes different labels for similar instances; (b)~a term that punishes misclassifications at the group-level. 
The general loss function $L(\boldsymbol{\theta})$ is then minimized as a function of the classifier parameters $\boldsymbol{\theta}$, 

\vspace{-.3cm}
\small
\begin{align}
L(\boldsymbol{\theta}) &= \frac{1}{N^2} \sum\limits_{i=1}^N \sum\limits_{j=1}^N \mathcal{S}(\boldsymbol{x}_i,\boldsymbol{x}_j) (y_i - y_j)^2 \nonumber\\
&+ \frac{\lambda}{K} \sum\limits_{k=1}^K (A(D_k,\boldsymbol{\theta}) - l_k)^2,  \label{eq:costgeneral}
\end{align}
\normalsize

where $\lambda$ is a free parameter that denotes the contribution of the group-level error to the loss function. In this function, $\mathcal{S}(\boldsymbol{x}_i,\boldsymbol{x}_j)$ measures the similarity between two instances $\boldsymbol{x}_i$ and $\boldsymbol{x}_j$, and $(y_i - y_j)^2$ denotes the square loss on the predictions for instances $i$ and $j$. In addition, $A(D_k,\boldsymbol{\theta})$ denotes the predicted label for the group $D_k$. Hence, the loss function punishes different labels for similar instances while still accounting for a correct classification of the groups.

In order to adapt the loss function to our problem, \ie classify sentences in financial news into positive and negative categories, we specify concrete functions for the placeholders in Equation \ref{eq:costgeneral} as follows. First, we use an rbf kernel to calculate a similarity measure between two sentence representations, \ie  $\mathcal{S}(\boldsymbol{x}_i,\boldsymbol{x}_j) = e^{-||\boldsymbol{x}_i - \boldsymbol{x}_j||^2} \in [0,1]$.  Second, we need to specify a classifier to predict $y_i$. Here, we choose a logistic regression model due to its simplicity and reliability. The prediction of the logistic regression model for the label of instance $i$ is given by $y_i = y_{\boldsymbol{\theta}}(\boldsymbol{x}_i) = \sigma(\boldsymbol{\theta}^T x_i)$ where $\sigma(x) = \frac{1}{1+ e^{-x}}$ denotes the value of the sigmoid function. 
Altogether, this results in a specific loss function which is to be minimized by the parameter of the logistic regression $\boldsymbol{\theta}$, 

\vspace{-.3cm}
\small
\begin{align} 
L(\boldsymbol\theta) &= \frac{1}{N^2} \sum\limits_{i=1}^N \sum\limits_{j=1}^N e^{-||\boldsymbol{x}_i - \boldsymbol{x}_j||^2} \left(\sigma(\boldsymbol{\theta}^T \boldsymbol{x}_i) - \sigma(\boldsymbol{\theta}^T \boldsymbol{x}_j)\right)^2   \nonumber \\
&+ \frac{\lambda}{K} \sum\limits_{k=1}^K\left(\frac{1}{|\mathcal{G}_k|} \left(\sum\limits _{x_i \in \mathcal{G}_k} \sigma(\boldsymbol{\theta}^T \boldsymbol{x}_i)\right) - l_k\right)^2. \label{eq:cost}
\end{align}
\normalsize

\begin{figure}[H]
  \centering
  \subfloat{\includegraphics[scale=0.16]{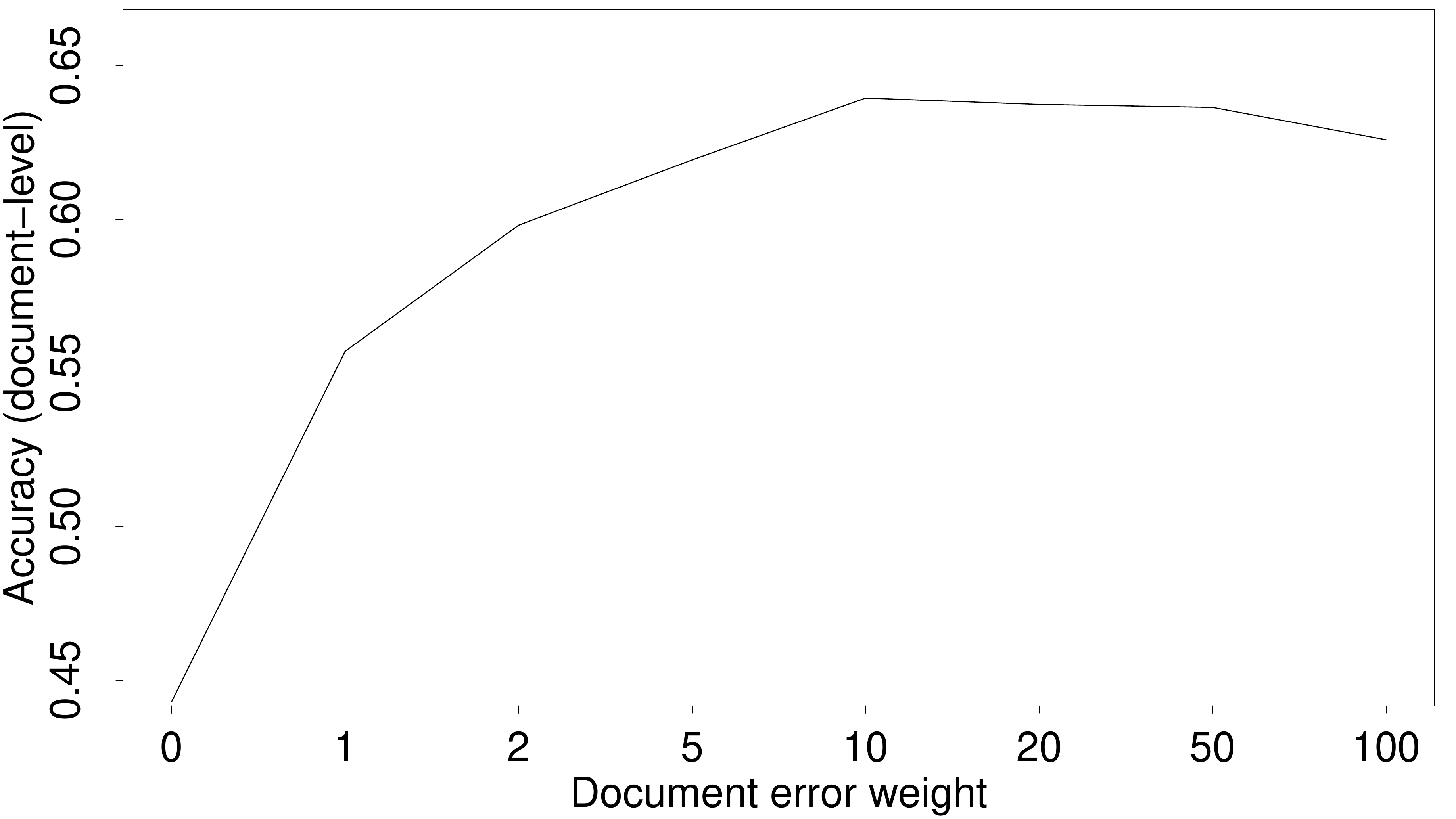}}
  \caption{Sensitivity for different values of document error weight parameter $\lambda$.}\label{fig:sensitivity}
\end{figure}

The parameter $\boldsymbol{\theta}$ is initialized with random values and optimized using stochastic gradient descent with momentum. In addition, we perform grid search to optimize the hyper parameters $\lambda$, learning rate, and momentum. According to our results, the model is most sensitive to changing the document error weight parameter $\lambda$, whereas learning rate and momentum have a smaller effect. The sensitivity for different values of $\lambda$ is also visualized in \Cref{fig:sensitivity}. 
Out of all considered models, we find the highest in-sample document-level accuracy of \num{64.40}\% using $\lambda=10$, learning rate = $0.05$, 
and momentum = $0.8$. 

Ultimately, we use the above model to predict labels of individual sentences as follows. First, a sentence is transformed into its vector representation $\boldsymbol{x}_i$. Second, we calculate $\sigma(\boldsymbol{\theta}^T \boldsymbol{x}_i)$ via the logistic regression model. If the result of $\sigma(\boldsymbol{\theta}^T \boldsymbol{x}_i)$ is greater than or equal to \SI{0.5}, the model predicts positive (and negative otherwise). The model is also capable of making predictions at the document-level. For this purpose, it chooses the most frequent label of all the sentences contained in the document, \ie positive documents are expected to contain a higher number of positive sentences than negative sentences and vice versa. 

\section{Evaluation}
\label{sec:results}

This section evaluates our method for inferring sentence-level sentiment in financial news. First, we present our model and illustrate an example of how our classifier can provide decision support for practitioners. Second, we compare the predictive performance of our method with several baseline approaches. Finally, we validate the robustness of our results using two additional datasets consisting of customer reviews. 

\subsection{Extraction of Sentence Labels}
\label{sec:sentenceLabels}

We use the methodology as described in the previous sections to infer sentence labels from ad~hoc announcements. 
The result of the learning procedure is a dataset containing documents that consist of groups of sentences in vector representations, where each sentence is assigned to a positive or negative polarity.

We proceed by presenting summary statistics of the resulting dataset. We find that a majority of \SI{53.16}{\percent} of all sentences are assigned a positive polarity, whereas the remaining \SI{46.84}{\percent} are assigned a negative polarity. \Cref{tbl:2x2} shows the number of occurrences of positive and negative sentences in our dataset, together with the resulting market reaction.
Specifically, we see that positive news contain \SI{57.70}{\percent} positive sentences and \SI{42.30}{\percent} negative sentences. In contrast, news with a negative market reaction contain only \SI{47.62}{\percent} positive sentences and \SI{52.38}{\percent} negative sentences. 

Interestingly, we observe that most ad~hoc announcements consist of a combination of positive and negative aspects. Specifically, out of all documents, \SI{97.57}{\percent} contain both positive and negative sentences. In addition, \SI{1.70}{\percent} of all documents contain only positive sentences, while \SI{0.73}{\percent} consist solely of negative sentences. We find two possible explanations for this overall high proportion of positive sentences: (1) the document labels feature a positive mean abnormal return, and (2) negative sentences in financial news typically exhibit greater length compared to positive sentences. 

\begin{table}
	\caption{Distribution of positive and negative sentences for different stock market reactions.}
	\centering
	\small
	\setlength{\extrarowheight}{3pt}
	\begin{tabular}{cc|c|c|}
		& \multicolumn{1}{c}{} & \multicolumn{2}{c}{Sentence label}\\
		& \multicolumn{1}{c}{} & \multicolumn{1}{c}{positive}  & \multicolumn{1}{c}{negative} \\\cline{3-4}
		\multirow{2}{*}{\rotatebox[origin=c]{90}{\parbox[c]{.9cm}{\centering Market reaction}}} & positive & \num{28926}~(\SI{57.70}{\percent}) & \num{21202}~(\SI{42.30}{\percent}) \\\cline{3-4}
		& negative & \num{19615}~(\SI{47.62}{\percent}) & \num{21572}~(\SI{52.38}{\percent}) \\\cline{3-4}
	\end{tabular}
	\label{tbl:2x2}
\end{table}

\subsection{Illustrative Example}
\label{sec:illustrative}

We now present an example of how our method for inferring sentence-level sentiment in financial news can provide decision support for practitioners, such as investors and investor relations departments. For this purpose, \Cref{fig:example} shows an excerpt of an ad~hoc announcement from the cable and harnessing manufacturing firm LEONI AG. This announcement was published on May 12, 2005 and led to an abnormal return of \SI{-4.6}{\percent} at the end of the trading day. The announcement consists of both positive and negative parts. While the positive parts describe increases in net income and margin expectations, the negative parts refer to lower expectations regarding future growth rates and the insolvency of a certain customer. 

According to \Cref{fig:example}, our classifier identifies all positive and negative parts correctly, including negated text fragments. Interestingly, applying traditional bag-of-words would be misleading in this case. For instance, because of a disregard for context, the first negative sentence would be classified positively, as it contains many positive words, such as \emph{\enquote{strong}}, \emph{\enquote{possible}}, and \emph{\enquote{growth}}. Overall, the example illustrates the challenges of accurate sentence classification in financial news. The identification of positive meaning is highly context-dependent and can result in entirely different interpretations when relying solely on word frequencies. As a remedy, our method can process complex sentences while preserving context and order of information. In addition, our model is solely trained on an objective response variable and thus adapts to domain-specific particularities of the given prose.

\begin{figure}
	\footnotesize
	\ttfamily
	[...]
	\ctextlightgray{
		As shown in LEONI AG's interim report, consolidated external sales amounted to EUR 350 million on 31 March 2005. The figure is therefore up about 23 percent on the same quarter one year earlier (EUR 284.8 million) despite difficult market conditions.}
	\ctextdarkgray{However, given the strong sales in quarters three and four of the previous year, it will not be possible to sustain this high rate of growth over 2005 as a whole.}
	\ctextlightgray{LEONI therefore reaffirms its sales forecast for fiscal 2005 of EUR 1.43 billion (up from EUR 1.25 billion in the previous year). Earnings before interest and taxes (EBIT) were up from EUR 9.5 million in the first quarter of 2004 to EUR 17.2 million in the same period this year, equating to growth of 81 percent. Consolidated net income increased by almost 44 percent, from EUR 5.5 million to EUR 7.9 million. In terms of operating earnings before interest and taxes (EBIT), the Company is a still aiming for a margin of seven percent over the year as a whole.}
	\ctextdarkgray{However, the insolvency of LEONI's customer MG Rover must be expected to incur exceptional charges of between five and seven million euros. It is not possible at this time to state the extent to which it might be possible to offset these charges during the current financial year.}
	[...]
	\normalfont
	\caption{Statistically positive and negative sentences in an exemplary ad~hoc announcement. Positive sentences are colored in light gray, whereas negative sentences are colored in dark gray.}
	\label{fig:example}
\end{figure}

\subsection{Predictive Performance on Manually-Labeled Sentences}
\label{sec:manual_label}

We now evaluate the predictive performance of our method on a manually-labeled dataset. For this purpose, we use a disjunct dataset that is labeled manually by three external persons with a background in finance. 

\begin{table*}
	\caption{Out-of-sample predictive performance. Left: Performance evaluation on manually-labeled sentences of financial news. Right: Predictive performance on document-level.}
	\sisetup{round-mode=places,round-precision=4}
	\sisetup{free-standing-units,table-align-text-post=false}
\renewcommand\arraystretch{0.8}
\centering
{ \setlength{\tabcolsep}{2pt}\footnotesize\begin{tabular}{l D{.}{.}{2.4} D{.}{.}{2.4}D{.}{.}{2.4}D{.}{.}{2.4}D{.}{.}{2.4} c D{.}{.}{2.4}  D{.}{.}{2.4}D{.}{.}{2.4}D{.}{.}{2.4}D{.}{.}{2.4}}
			\toprule
			&\multicolumn{5}{c}{\textbf{\begin{tabular}{c}Evaluation: Sentence-Level\end{tabular}}} && \multicolumn{5}{c}{\textbf{\begin{tabular}{c}Evaluation: Document-Level\end{tabular}}} \\
			\cmidrule{2-6} \cmidrule{8-12}
			\textbf{Method}  & \multicolumn{1}{c}{\textbf{\begin{tabular}{c}Accuracy\end{tabular}}} & \multicolumn{1}{c}{\textbf{\begin{tabular}{c}Recall\end{tabular}}} & \multicolumn{1}{c}{\textbf{\begin{tabular}{c}Precision\end{tabular}}} & \multicolumn{1}{c}{\textbf{\begin{tabular}{c}$\boldsymbol{F_1}$-Score\end{tabular}}}  & \multicolumn{1}{c}{\textbf{\begin{tabular}{c}Neutral\end{tabular}}} 
			 && \multicolumn{1}{c}{\textbf{\begin{tabular}{c}Accuracy\end{tabular}}} & \multicolumn{1}{c}{\textbf{\begin{tabular}{c}Recall\end{tabular}}} & \multicolumn{1}{c}{\textbf{\begin{tabular}{c}Precision\end{tabular}}} & \multicolumn{1}{c}{\textbf{\begin{tabular}{c}$\boldsymbol{F_1}$-Score\end{tabular}}}  & \multicolumn{1}{c}{\textbf{\begin{tabular}{c}Neutral\end{tabular}}}\\
			\midrule
			\textsc{\underline{Dictionaries}} \\
			Harvard IV & 48.00 \% & 75.33 \% & 48.71 \% & 59.16 \% & 22.67 \% && 50.00 \% & 99.64 \% & 50.00 \% & 66.59 \% & 0.36 \%  \\ 
			Loughran-McDonald & 31.67 \% & 25.33 \% & 29.00 \% & 27.05 \% & 53.00 \%  && 51.92 \% & 39.78 \% & 52.53 \% & 45.28 \% & 9.31 \% \\
			\addlinespace
			\textsc{\underline{Bag-of-words}} \\
			Logistic regression  & 55.40 \% & 60.40 \% & 54.91 \% & 57.52 \% & \NEUTRAL && 53.38 \% & 45.26 \% & 54.03 \% & 49.26 \% &\NEUTRAL \\
			Random forest & 54.60 \% & 96.40 \% & 52.51 \% & 67.98 \% &\NEUTRAL && 55.29 \% & 75.36 \% & 53.78 \% & 62.77 \%  & \NEUTRAL  \\
			Support vector machine & 56.40 \% & 63.00 \% & 55.65 \% & 59.10 \%  & \NEUTRAL &&  53.28 \% & 63.87 \% & 52.71 \% & 57.76  \% &\NEUTRAL\\
			Artificial Neural Network & 58.30 \% & 55.80 \% & 58.74 \% & 57.23 \% & \NEUTRAL && 54.20 \% & 61.50 \% & 53.66 \% & 57.31 \% &\NEUTRAL\\
			\addlinespace
			\textsc{\underline{Sentence embeddings}} \\
			Logistic regression & 64.90 \% & 76.80 \% & 62.04 \% & 68.63 \% & \NEUTRAL  & & 57.85 \% & 58.58 \% & 57.74 \% & 58.15 \% & \NEUTRAL  \\
			Random forest & 61.60 \% & 81.00 \% & 58.36\% & 67.84 \% & \NEUTRAL & & 56.39 \% & 82.85 \% & 54.18 \% & 65.51 \% & \NEUTRAL\\
			Support vector machine & 65.60 \% & 65.80 \% & 65.54 \% & 65.66 \% & \NEUTRAL  && 56.85 \% & 58.21 \% & 56.66 \% & 57.43 \%   &\NEUTRAL \\
			Artificial Neural Network & 66.10 \% & 75.80 \% & 63.48 \% & 69.10 \% & \NEUTRAL && 57.21 \% & 64.23 \% & 56.32 \% & 60.02 \% &\NEUTRAL \\
			\addlinespace
			Our approach (MIL) & 69.90 \% & 67.80 \% & 70.77 \% & 69.25 \%  &\NEUTRAL & & 55.84 \% & 67.36 \% & 54.75 \% & 60.39 \% & \NEUTRAL\\
			\bottomrule
	\end{tabular}}
	\label{tbl:results_sentence_level}
\end{table*}

The dataset consists of \num{1000} randomly drawn sentences from ad~hoc announcements, with an equal number of \num{500} positive and \num{500} negative sentences\footnote{Our dataset is available from \url{https://github.com/InformationSystemsFreiburg/SentenceLevelSentimentFinancialNews.}}. 
We use this dataset to compare the predictive performance of our approach to several baseline methods. First, we employ common sentiment dictionaries for polarity detection, namely the Harvard IV dictionary \cite{Stone.2002} and the Loughran-McDonald dictionary \cite{Loughran.2011}, the latter of which was developed for finance-specific texts. These dictionaries are a frequent choice when it comes to sentiment analysis of financial news (\eg~\cite{Tetlock.2008, Garcia.2013}). Second, we employ the bag-of-words approach in combination with common machine learning classifiers for text categorization, \ie logistic regression, random forest, support vector machine and artificial neural network\footnote{We optimize the hyperparameters of the machine learning classifiers using grid search based on 5-fold cross-validation.}. Third, we train the machine learning models based on sentence embeddings. We train all of these models on the dataset that is used in the previous sections. 

The left panel in \Cref{tbl:results_sentence_level} compares the predictive performance of our approach with the baseline methods. Our approach yields an accuracy of \SI{69.90}{\percent} on the manually-labeled dataset. This is at least \num{3.80} percentage points higher than the best-performing baseline method, \ie the artifical neural network trained on sentence embeddings. We also see that all machine learning models yield a higher predictive performance when being trained on sentence embeddings instead of bag-of-words feature representations. In addition, we note that the frequently-employed dictionaries are not suitable for sentence-level sentiment analysis of financial news. In fact, \Cref{tbl:results_sentence_level} reveals that the Harvard IV dictionary classifies \SI{22.67}{\percent} of all sentences as neutral. We observe a similar pattern for the finance-specific Loughran-McDonald dictionary, which assigns \SI{53.00}{\percent} of all sentences to a neutral class. There are two reasons for this result: first, dictionary-based approaches predict a neutral class if the number of positive polarity words equals the number of negative polarity words. Second, the polarity dictionary does not contain any of the words in a given sentence. 

\subsection{Predictive Performance on Document-Level}

Next, we evaluate the performance of our model as a document-level classifier. For this purpose, we compare the document-level predictions of our method with the document labels, \ie the abnormal returns. In a first step, we split our dataset of ad~hoc announcements in an 80:20 ratio for training and testing, so that the announcements of the training set are older than the announcements in the test set. This procedure precludes learning anomalies based on information which would only be available ex-post \cite{Kraus.2017}. Subsequently, we compare the results of our method with the same baseline classifiers from the previous sections, \ie dictionary-based approaches and machine-learning methods. 

\begin{table*}
\caption{Out-of-sample predictive performance for customer reviews with sentence-level annotations.}
	\sisetup{round-mode=places,round-precision=4}
	\sisetup{free-standing-units,table-align-text-post=false}
\renewcommand\arraystretch{0.8}
\centering
{ \setlength{\tabcolsep}{2pt}\footnotesize\begin{tabular}{l D{.}{.}{2.4} D{.}{.}{2.4}D{.}{.}{2.4}D{.}{.}{2.4}D{.}{.}{2.4} c D{.}{.}{2.4}  D{.}{.}{2.4}D{.}{.}{2.4}D{.}{.}{2.4}D{.}{.}{2.4}}
			\toprule
			&\multicolumn{5}{c}{\textbf{\begin{tabular}{c}\textsc{Study I: IMDb movie reviews}\end{tabular}}} && \multicolumn{5}{c}{\textbf{\begin{tabular}{c}\textsc{Study II: Yelp restaurant Reviews}\end{tabular}}} \\
			\cmidrule{2-6} \cmidrule{8-12}
			\textbf{Method}  & \multicolumn{1}{c}{\textbf{\begin{tabular}{c}Accuracy\end{tabular}}} & \multicolumn{1}{c}{\textbf{\begin{tabular}{c}Recall\end{tabular}}} & \multicolumn{1}{c}{\textbf{\begin{tabular}{c}Precision\end{tabular}}} & \multicolumn{1}{c}{\textbf{\begin{tabular}{c}$\boldsymbol{F_1}$-Score\end{tabular}}}  & \multicolumn{1}{c}{\textbf{\begin{tabular}{c}Neutral\end{tabular}}} 
			 && \multicolumn{1}{c}{\textbf{\begin{tabular}{c}Accuracy\end{tabular}}} & \multicolumn{1}{c}{\textbf{\begin{tabular}{c}Recall\end{tabular}}} & \multicolumn{1}{c}{\textbf{\begin{tabular}{c}Precision\end{tabular}}} & \multicolumn{1}{c}{\textbf{\begin{tabular}{c}$\boldsymbol{F_1}$-Score\end{tabular}}}  & \multicolumn{1}{c}{\textbf{\begin{tabular}{c}Neutral\end{tabular}}}\\
			\midrule
			\textsc{\underline{Dictionaries}} \\
			Harvard IV & 60.30 \% & 74.20 \% & 58.06 \% & 65.14 \% & 22.90 \%  & & 53.60 \% & 70.60 \% & 52.69 \% & 60.34 \% & 24.50 \% \\
			Loughran-McDonald & 38.40 \% & 35.80 \% & 37.76 \% & 36.76 \% & 51.30 \% && 37.70 \% & 43.60 \% & 39.00 \% & 41.17 \% & 52.20 \% \\
			\addlinespace
			\textsc{\underline{Bag-of-words}} \\
			Logistic regression & 83.40 \% & 82.00 \% & 84.36 \% & 83.16 \% & \NEUTRAL& & 83.80 \% & 83.80 \% & 83.80 \% & 83.80 \% & \NEUTRAL \\
			Random forest & 69.70 \% & 98.20 \% &  62.55 \% & 76.42 \% &\NEUTRAL && 80.50 \% & 89.60 \% & 75.80 \% & 82.13 \% &\NEUTRAL\\
			Support vector machine & 78.70 \% & 92.40 \% & 72.53 \% & 81.27 \% & \NEUTRAL && 84.50 \%  & 86.20 \% & 83.37 \% & 84.76 \% &\NEUTRAL \\
			Artificial Neural Network & 80.80 \% & 84.60 \% & 78.62 \% & 81.50 \% & \NEUTRAL && 83.20 \% & 79.40 \% & 85.93 \% & 82.54 \% & \NEUTRAL \\
			\addlinespace
			\textsc{\underline{Sentence embeddings}} \\
			Logistic regression & 84.50 \% & 83.00 \% & 85.57 \% & 84.27 \% & \NEUTRAL& & 85.40 \% & 85.80 \% & 85.12 \% & 85.49 \% &\NEUTRAL \\
			Random forest & 77.60 \% & 80.80 \% & 75.94 \% & 78.29 \%  & \NEUTRAL& & 80.90 \% & 77.00 \% & 83.51 \% & 80.12 \%  &\NEUTRAL\\
			Support vector machine & 85.20 \%  & 85.40 \% & 85.06 \% & 85.23 \% &\NEUTRAL& & 85.10 \% & 83.60 \% & 86.19 \% & 84.87 \%   &\NEUTRAL\\
			Artificial Neural Network & 84.00 \% & 84.80 \% & 83.46 \% & 84.12 \% & \NEUTRAL && 84.80 \% & 83.60 \% & 85.66 \% & 84.62 \% & \NEUTRAL \\
			\addlinespace
			Our approach (MIL) & 86.40 \% & 85.60 \% & 83.92 \% & 84.75 \% &\NEUTRAL  &&  86.30 \% &  85.60 \% & 86.82 \% & 86.20 \% &\NEUTRAL\\ 	
			\bottomrule
	\end{tabular}}
\label{tbl:results_machine_learning}
\end{table*}

The results are shown in the right panel of \Cref{tbl:results_sentence_level}. According to our results, our approach yields a document-level accuracy of \SI{55.84}{\percent} on out-of-sample documents. This is only \num{2.01} percentage points lower compared to the best performing baseline method (logistic regression) for document-level text classification. 
As a result, our approach presents a viable alternative that competes well with traditional machine learning models at the document-level but, at the same time, guarantees full interpretability at the sentence-level. Moreover, we see that the method is capable of successfully transferring information from the document-level to the sentence-level, and back again from sentences to documents.

\subsection{Robustness Check Using Customer Reviews}

Finally, we validate the benefits of our model for other text sources. For this purpose, we utilize two additional datasets from the related literature, namely \num{25000} IMDb movie reviews\footnote{Available from {\url{http://ai.stanford.edu/~amaas/data/sentiment/}}.} and \num{60000} Yelp restaurant reviews\footnote{Available from {\url{https://www.yelp.com/dataset/challenge}}}. Both datasets contain an equal number of positive and negative reviews, where each review is annotated with an overall rating at the document-level. In addition, we use the datasets created by \cite{Kotzias.2015} that contain a balanced number of manually-labeled sentences. We thus train individual models for both datasets using the same methodology as described in the previous sections. \Cref{tbl:results_machine_learning} compares the sentence-level predictive performance of our approach with the baseline methods. In the case of IMDb movie reviews (Study I), our approach yields a predictive accuracy of to \SI{86.40}{\percent}, which outperforms the traditional models, as well as dictionary-based approaches, by at least \num{1.20} percentage points. We observe a similar pattern for the Yelp restaurant reviews (Study II). Here our method yields a predictive accuracy of up to \SI{86.30}{\percent}, exceeding the performance of alternative approaches by at least {\num{0.90}} percentage points. Overall, this shows that the method is not limited to finance-related texts but also a highly interesting tool for text classification applications in other domains, such as marketing.

\section{Discussion}
\label{sec:discussion}

Our study not only allows for a better comprehension of decision-making in a financial context, but is also highly relevant for communication professionals and investors. 

First and foremost, this work entails multiple implications for possible enhancements of methods for sentiment analysis of financial news. It shows that current sentiment analysis approaches are not adequate for studying the reception of financial news on a fine-granular level. Corresponding inferences for individual sentences result in low explanatory power and lower predictive performance. This also coincides with \cite{Li.2010b}, who suggest that the \emph{\enquote{dictionary approach might not work well for analyzing the tone of corporate filings.}} Moreover, we see that machine learning algorithms, ignoring the characteristics of multi-instance problems, perform worse in this scenario \cite{Dietterich.1997}. As a remedy, we propose the use of distributed text representations and multi-instance learning to infer sentences with a positive or negative polarity. By incorporating context and domain-specific features, this methodology can be used to study the reception of individual text fragments in presence of a document label, such as stock market returns. 
 Future research can thus benefit from a method that uses statistical rigor to study the reception of financial news on a fine-grained level without the need for any kind of manual labeling.
Yet, the proposed method is not limited to the study of sentence-level sentiment in financial news. In fact, one can easily adapt it to
all applications of natural language processing which utilize a decision variable and where the information can be separated into different subgroups, such as
sentences or paragraphs. 


This paper also provides managerial decision support for companies by addressing the question of how individual text components in their corporate disclosures are actually perceived by investors. In a next step, managers and investor relations departments can benefit from a self-reflective writing process that avoids noisy signals in their communications, thus helping to prevent stock prices from deviating from the expected value. In a similar vein, they can use our method for inferring sentence-level sentiment to analyze the performance of their past disclosures and to monitor the form and style relative to their competitors. 


Ultimately, the presented approach can provide decision support for news-driven trading. In this context, we present an intriguing tool to practitioners for the purpose of improving the automated processing of financial news in their information systems. For example, our approach can be integrated into graphical tools that are targeted to financial professionals or private traders seeking to process large quantities of disclosures. Among others advantages, such tools would be able to assist traders in processing financial information by highlighting relevant positive and negative text fragments. Overall, our methodology can enhance the accuracy of decision support based on textual data and can be seamlessly integrated into an existing tool chain.

\section{Conclusion}
\label{sec:conclusion}


Automated decision support for financial news requires robust methods which operationalize the reception of texts on a fine-grained level. For this purpose, this paper proposes the use of distributed text representations and multi-instance learning to analyze the sentiment of individual sentences in financial news with high interpretability. In contrast to previous approaches that merely predict the stock market reaction in response to news items on a document-level, our method transfers information from the document-level to the sentence-level. According to our results, the proposed approach outperforms existing methods by at least \num{3.80} percentage points on a manually labeled dataset of sentences of financial news. 


Our study immediately suggests manifold implications for researchers and practitioners. Financial professionals and investors can benefit from our method, which allows them to easily distinguish between positive and negative text fragments in financial news based on statistical rigor. In addition, company executives and investor relations departments may wish to consider choosing their language strategically to ensure that their message is interpreted as intended. Ultimately, it is hoped that the datasets and method presented in this paper will be used in future research in order to yield novel insights into behavioral and finance research questions.

In future work, we will advance our study as follows: first, from a methodological point of view, the application of multi-instance learning is not restricted to logistic regression. Although a comparison to alternative classifiers is beyond the scope of this paper, we expect other sophisticated models to achieve similar performance on the utilized datasets. In addition, the implementation of alternative loss functions might provide an avenue to further improve the predictive performance. Second, our method for inferring fine-grained sentiment labels for individual sentences also serves as a powerful tool to assess the effects of narrative impression management techniques on the perception of investors and to infer behavioral implications. Corresponding research questions have been difficult or impossible to analyze in previous works since the nature of language provides countless possibilities to express the same meaning in different words. Third, further research is necessary to study the differences in information reception among different target groups. For instance, people might interpret news differently depending on their information processing skills and the subjective interpretation of the same information might vary across different audiences and cultures.


\bibliographystyle{ieeetr}
\bibliography{bib/literature}

\begin{thebibliography}{10}

\bibitem{Benston.1973}
G.~J. Benston, ``Required disclosure and the stock market: An evaluation of the
  securities exchange act of 1934,'' {\em The American Economic Review},
  vol.~63, no.~1, pp.~132--155, 1973.

\bibitem{Kearney.2014}
C.~Kearney and S.~Liu, ``Textual sentiment in finance: A survey of methods and
  models,'' {\em International Review of Financial Analysis}, vol.~33, no.~1,
  pp.~171--185, 2014.

\bibitem{Loughran.2016}
T.~Loughran and B.~McDonald, ``Textual analysis in accounting and finance: A
  survey,'' {\em Journal of Accounting Research}, vol.~54, no.~4,
  pp.~1187--1230, 2016.

\bibitem{Tetlock.2008}
P.~C. Tetlock, M.~Saar-Tsechansky, and S.~Macskassy, ``More than words:
  Quantifying language to measure firms' fundamentals,'' {\em The Journal of
  Finance}, vol.~63, no.~3, pp.~1437--1467, 2008.

\bibitem{Loughran.2011}
T.~Loughran and B.~McDonald, ``When is a liability not a liability? textual
  analysis, dictionaries, and 10-ks,'' {\em The Journal of Finance}, vol.~66,
  no.~1, pp.~35--65, 2011.

\bibitem{Carter.1999}
M.~E. Carter and B.~S. Soo, ``The relevance of form 8-k reports,'' {\em Journal
  of Accounting Research}, vol.~37, no.~1, pp.~119--132, 1999.

\bibitem{Alfano.2018}
S.~Alfano, N.~Pr{\"o}llochs, S.~Feuerriegel, and D.~Neumann, ``Say it right: Is
  prototype to enable evidence-based communication using big data,'' in {\em
  Analytics and Data Science}, pp.~217--221, Springer, 2018.

\bibitem{Prollochs.2016b}
N.~Pr{\"o}llochs, S.~Feuerriegel, and D.~Neumann, ``Negation scope detection in
  sentiment analysis: Decision support for news-driven trading,'' {\em Decision
  Support Systems}, vol.~88, pp.~67--75, 2016.

\bibitem{Schumaker.2012}
R.~P. Schumaker, Y.~Zhang, C.-N. Huang, and H.~Chen, ``Evaluating sentiment in
  financial news articles,'' {\em Decision Support Systems}, vol.~53, no.~3,
  pp.~458--464, 2012.

\bibitem{Prollochs.2015}
N.~Pr{\"o}llochs, S.~Feuerriegel, and D.~Neumann, ``Generating domain-specific
  dictionaries using bayesian learning,'' in {\em Proceedings of the 23rd
  European Conference on Information Systems (ECIS)}, (M{\"u}nster, Germany),
  2015.

\bibitem{Lerman.2010}
A.~Lerman and J.~Livnat, ``The new form 8-k disclosures,'' {\em Review of
  Accounting Studies}, vol.~15, no.~4, pp.~752--778, 2010.

\bibitem{Abbasi.2016}
A.~Abbasi, S.~Sarker, and R.~H. Chiang, ``Big data research in information
  systems: Toward an inclusive research agenda,'' {\em Journal of the
  Association for Information Systems}, vol.~17, no.~2, p.~3, 2016.

\bibitem{Kotzias.2015}
D.~Kotzias, M.~Denil, N.~de~Freitas, and P.~Smyth, ``From group to individual
  labels using deep features,'' in {\em Proceedings of the 21st International
  Conference on Knowledge Discovery and Data Mining (SIGKDD)}, pp.~597--606,
  2015.

\bibitem{GroKlumann.2011}
A.~Gro{\ss}-Klu{\ss}mann and N.~Hautsch, ``When machines read the news: Using
  automated text analytics to quantify high frequency news-implied market
  reactions,'' {\em Journal of Empirical Finance}, vol.~18, no.~2,
  pp.~321--340, 2011.

\bibitem{Hannemann.2018}
F.~Hannemann, N.~Pr{\"o}llochs, and D.~Neumann, ``Noise trader behavior -- a
  disaggregated approach to news reception and processing in financial
  markets,'' in {\em Proceedings of the 26th European Conference on Information
  Systems (ECIS)}, 2018.

\bibitem{Prollochs.2018}
N.~Pr{\"o}llochs and S.~Feuerriegel, ``Business analytics for strategic
  management: Identifying and assessing corporate challenges via topic
  modeling,'' {\em Information {\&} Management}, vol.~forthcoming, 2018.

\bibitem{Tetlock.2007}
P.~C. Tetlock, ``Giving content to investor sentiment: The role of media in the
  stock market,'' {\em The Journal of Finance}, vol.~62, no.~3, pp.~1139--1168,
  2007.

\bibitem{Henry.2008}
E.~Henry, ``Are investors influenced by how earnings press releases are
  written?,'' {\em Journal of Business Communication}, vol.~45, no.~4,
  pp.~363--407, 2008.

\bibitem{Nassirtoussi.2014}
A.~K. Nassirtoussi, S.~Aghabozorgi, T.~Y. Wah, and D.~C.~L. Ngo, ``Text mining
  for market prediction: A systematic review,'' {\em Expert Systems with
  Applications}, vol.~41, no.~16, pp.~7653--7670, 2014.

\bibitem{Chiong.2018}
R.~Chiong, Z.~Fan, Z.~Hu, M.~T.~P. Adam, B.~Lutz, and D.~Neumann, ``A sentiment
  analysis-based machine learning approach for financial market prediction via
  news disclosures,'' in {\em Proceedings of the Genetic and Evolutionary
  Computation Conference Companion}, pp.~278--279, 2018.

\bibitem{Kraus.2017}
M.~Kraus and S.~Feuerriegel, ``Decision support from financial disclosures with
  deep neural networks and transfer learning,'' {\em Decision Support Systems},
  no.~104, pp.~38--48, 2017.

\bibitem{Allee.2015}
K.~D. Allee and M.~D. Deangelis, ``The structure of voluntary disclosure
  narratives: Evidence from tone dispersion,'' {\em Journal of Accounting
  Research}, vol.~53, no.~2, pp.~241--274, 2015.

\bibitem{Li.2010b}
F.~Li, ``Textual analysis of corporate disclosures: A survey of the
  literature,'' {\em Journal of Accounting Literature}, vol.~29, p.~143, 2010.

\bibitem{Cortis.2017}
K.~Cortis, A.~Freitas, T.~Daudert, M.~Huerlimann, M.~Zarrouk, S.~Handschuh, and
  B.~Davis, ``Fine-grained sentiment analysis on financial microblogs and
  news,'' in {\em Proceedings of the 11th International Workshop on Semantic
  Evaluation (SemEval)}, pp.~519--535, 2017.

\bibitem{Carbonneau.2018}
M.-A. Carbonneau, V.~Cheplygina, E.~Granger, and G.~Gagnon, ``Multiple instance
  learning: A survey of problem characteristics and applications,'' {\em
  Pattern Recognition}, vol.~77, pp.~329--353, 2018.

\bibitem{Dundar.2007}
M.~Dundar, B.~Krishnapuram, R.~B. Rao, and G.~M. Fung, ``Multiple instance
  learning for computer aided diagnosis,'' in {\em Proceedings of the 20th
  Neural Information Processing Systems Conference (NIPS)}, pp.~425--432, 2007.

\bibitem{Groth.2011}
S.~S. Groth and J.~Muntermann, ``An intraday market risk management approach
  based on textual analysis,'' {\em Decision Support Systems}, vol.~50, no.~4,
  pp.~680--691, 2011.

\bibitem{Hagenau.2013}
M.~Hagenau, M.~Liebmann, and D.~Neumann, ``Automated news reading: Stock price
  prediction based on financial news using context-capturing features,'' {\em
  Decision Support Systems}, vol.~55, no.~3, pp.~685--697, 2013.

\bibitem{Prollochs.2018b}
N.~Pr{\"o}llochs, M.~Adam, S.~Feuerriegel, and D.~Neumann, ``Information
  processing of financial news: The role of cognitive dissonance and
  information avoidance,'' in {\em Proceedings of the 26th European Conference
  on Information Systems (ECIS)}, 2018.

\bibitem{MacKinlay.1997}
A.~C. MacKinlay, ``Event studies in economics and finance,'' {\em Journal of
  Economic Literature}, vol.~35, no.~1, pp.~13--39, 1997.

\bibitem{Zhang.2012}
Y.~Zhang, P.~E. Swanson, and W.~Prombutr, ``Measuring effects on stock returns
  of sentiment indexes created from stock message boards,'' {\em Journal of
  Financial Research}, vol.~35, no.~1, pp.~79--114, 2012.

\bibitem{Manning.2014}
C.~D. Manning, M.~Surdeanu, J.~Bauer, J.~Finkel, S.~J. Bethard, and
  D.~McClosky, ``The stanford corenlp natural language processing toolkit,'' in
  {\em Proceedings of the 52nd Annual Meeting of the Association for
  Computational Linguistics}, pp.~55--60, 2014.

\bibitem{Le.2014}
Q.~Le and T.~Mikolov, ``Distributed representations of sentences and
  documents,'' in {\em Proceedings of the 31st International Conference on
  Machine Learning (ICML)}, pp.~1188--1196, 2014.

\bibitem{Tang.2014b}
T.~Tang, E.~Fang, and F.~Wang, ``Is neutral really neutral? the effects of
  neutral user-generated content on product sales,'' {\em Journal of
  Marketing}, vol.~78, no.~4, pp.~41--58, 2014.

\bibitem{Lau.2016}
J.~H. Lau and T.~Baldwin, ``An empirical evaluation of doc2vec with practical
  insights into document embedding generation,'' in {\em Proceedings of the 1st
  Workshop on Representation Learning for NLP}, pp.~78--86, 2016.

\bibitem{Dietterich.1997}
T.~G. Dietterich, R.~H. Lathrop, and T.~Lozano-P{\'e}rez, ``Solving the
  multiple instance problem with axis-parallel rectangles,'' {\em Artificial
  Intelligence}, vol.~89, no.~1, pp.~31--71, 1997.

\bibitem{Stone.2002}
P.~J. Stone, ``General inquirer harvard-iv dictionary,'' 2002.

\bibitem{Garcia.2013}
D.~Garcia, ``Sentiment during recessions,'' {\em The Journal of Finance},
  vol.~68, no.~3, pp.~1267--1300, 2013.

\end{thebibliography}
\end{document}